\documentclass[journal]{IEEEtran}

\usepackage{balance}
\usepackage{comment}
\usepackage{subfigure}
\usepackage{graphicx}
\usepackage{epsfig}
\usepackage{url}
\usepackage{amsfonts}
\usepackage{amssymb}
\usepackage{mdwmath}
\usepackage{mdwtab}
\usepackage{cite}
\usepackage{booktabs}
\usepackage{tabularx}
\usepackage{multirow} 
\usepackage[]{caption}
\usepackage{amsmath}
\usepackage{graphicx}
\usepackage{hyperref}
\usepackage{xcolor}

\begin{document}

\title{Towards a Practical Architecture for India Centric Internet of Things}
\author{\IEEEauthorblockN{Prasant Misra$^\ddagger$, Yogesh Simmhan$^{\dagger}$, Jay Warrior$^\ddagger$}\\
\small{\textit{(contributors in alphabetical order)}}\\
\IEEEauthorblockA{$^{\ddagger}$Robert Bosch Centre for Cyber Physical Systems (RBCCPS), www.rbccps.org\\
									$^{\dagger}$DREAM:Lab, Supercomputer Education and Research Centre (SERC), www.dream-lab.serc.iisc.ernet.in\\			
									Indian Institute of Science, Bangalore, India\\
Email: \{prasant.misra, jay.warrior\}@rbccps.org$\ddagger$ simmhan@serc.iisc.in$^{\dagger}}$
}

\maketitle

\begin{abstract}
An effective architecture for the Internet of Things (IoT), particularly
for an emerging nation like India with limited technology penetration at the
national scale, should be based on tangible technology advances in the
present, practical application scenarios of social and entrepreneurial
value, and ubiquitous capabilities that make the realization of IoT
affordable and sustainable. Humans, data, communication and devices play key
roles in the IoT ecosystem that we perceive. In a push towards this sustainable
and practical IoT Architecture for India, we synthesize ten design
paradigms to consider.
\end{abstract}

\section{Introduction}

The current, widespread thinking on the Internet of Things (IoT)\cite{IoT-History, eu-iot, cisco-iot, ibm-iot, medcitynews, googlenest, retail, amazon-shipping, iot-china} makes several, arguably misplaced assumptions, driven by repackaged products or completely clean-slate (and costly) designs. Some of these assumptions for an IoT architecture are:
\begin{itemize}
\item Hundreds of devices as part of a tightly coupled infrastructure
\item Devices costing \$5-500, and are customized for a specific IoT application
\item Structured communication networks (IPv6) with always-on connectivity
\item Cloud-centric data collection and analysis, with centralized control
\item Single vendor owns the platform, Cloud services, data and eco-system for an application
\end{itemize}
Examples of these vertically integrated IoT silos include smart power meters and SCADA systems, personal monitoring devices like FitBit, etc.,.
However, an effective architecture for IoT, particularly for an emerging nation like India\cite{yourstory} with limited technology penetration at the national scale, should be based on: ($1$) tangible technology advances in the present, ($2$) practical application scenarios of social and entrepreneurial value, and ($3$) ubiquitous capabilities that make the realization of IoT affordable and sustainable. 
A rethink of the above assumptions would give us:
\begin{itemize}
\item Thousands of loosely connected devices in immediate vicinity, and millions more further out
\item Devices that cost from \$0.01-\$3 combined with existing in-person generic devices like smart phones
\item A mix of ad hoc P2P, 2G/3G/4G and WiFi based on existing and emerging communication networks, and with intermittent connectivity
\item Data collection and personalized analytics that seamlessly span edge devices and the Cloud, with control over data sharing and ownership while encouraging Open Data
\item Open ecosystem without vendor lock-in using standard Internet and Web protocols, allowing devices and data to be shared across IoT applications
\end{itemize}

\section{Design Paradigms}

Humans, data, communication and devices play key roles in the IoT ecosystem.
In a push towards this sustainable and practical IoT architecture for India, we synthesize ten design paradigms to consider. 
\vspace{-3mm}
\subsection{Human-centric rather than Thing-centric}

Current IoT architectures are device or network oriented. However, the key value proposition of IoT is from the interaction of these ``Things'' with humans and society, and the benefits gained for humans who are part of, affected by and influence the network. Technologies, services and decision making must create an IoT experience that deeply engages with people. This may be mundane, like providing optimal traffic routing in a smart city, or essential, like offering personalized health suggestions for patients. As a result, devices, networks, data, and analytics that are in close proximity to humans and widely prevalent – smart and feature phones, wireless interfaces such as Bluetooth/WiFi/$2$G/$3$G, wearables/body area networks, etc., – find favour.

\subsection{Span Virtual and Physical Worlds}

Much of the IoT conversation is about the physical infrastructure and its optimization. Bringing in the human and a social element, with their virtual online avatars (social networks and virtual agents), helps span the digital and physical world, and across humans and infrastructure. Capturing proximity and interactions between humans and ``Things'' (H$2$H, H$2$M, M$2$M) in the physical and virtual worlds are necessary for actionable intelligence. 

\subsection{Big-Little Data}

Analytics performed on information from diverse sources within the IoT architecture helps with data-driven decision making. Two classes of data are transient sensor and personal data collected continuously from a humans/physical devices ``Little'' data) and persistent knowledge-bases and archives that span domains available in central repositories/Clouds (``Big'' data). Meaningful analytics requires both ``Big'' and ``Little'' data to be combined, and often in real time.

\subsection{Analytics from the Edge to the Cloud}

Related to ``Big-Little'' data is performing distributed analytics and decision making. The current model of pushing all data to a central Cloud for analytics will not scale, is inefficient, and raises privacy concerns. Given the enhanced capabilities of edge devices like smart phones and potentially intermittent communication, we need to decide if/when to push a subset of the ``Big'' data to the phone, if/when to push a subset of the ``Little'' data to the Cloud, and how to make collaborative decisions automatically. Communication and compute capability, data privacy and availability, and end-use application inform these choices.

\subsection{Bring the Network to the Sensor}

As we witness an immense proliferation of tens of thousands of cheap IoT devices, these will be constrained in energy and communication capabilities. Rather than rely on massive deployment of custom sensor networks and new standards, there is more value in piggybacking on existing, widely adopted standards and reusing symbiotic infrastructure. Using phones as P$2$P data mules for last mile connectivity, combined with highly functional gateways and Clouds for coordination suggests an asymmetric architecture.

\subsection{How ``Low'' can you go ?}

Technology penetration has not been uniform across countries, regions, or for that matter, industries.
This disparity is a reflection of the differences in infrastructure, cost of access, telecommunication networks and services, policies, etc., among different economies.
Therefore, the cost and technology behind the sensing, device, networking and analytic solutions for the IoT should be affordable and scale to billions of users.
In this regard, reuse of commodity hardware and existing infrastructure would be critical to the IoT success.
Hence, reusable devices and sensors used in novel ways are preferable over custom solutions with cutting-edge capabilities, or canned solutions developed for advanced economies.

\subsection{Whose data is it anyway ?}

IoT offers immense opportunities for innovation and entrepreneurship. 
The intersection of devices, communication, data and humans offers interesting incentive and business models. 
A key success of the WWW is the ability for businesses to monetize users' data e.g., Google Ad revenue using user's personal data in return for free search and mail services to end users. 
With IoT, devices are going to be even more closer to humans and blend into our environment. 
Ensuring there is transparency in data ownership, sharing, and usage is important. 
Further, there is scope for data brokering that encourages open data sharing by users with business in return for clear rewards, be they monetary, peer recognition, or the greater good.

\subsection{When ``good enough'' is enough ?}

IoT is naturally a diverse ecosystem with unreliability and uncertainties built in. 
Cheap sensors mean questionable data quality. 
Humans are fickle to model and even physical systems complex. 
Distributed thing and intermittent communication are a given. 
Data privacy puts bounds on its availability. 
As a result, analytic and decision making have to be probabilistic and the system and application has to be conscious of what is ``good enough'', and not fail in the absence of perfect behavior. 

\subsection{Context determines the Action}

Given the uncertainties of the system and humans being central entities, much of the decision making within the IoT infrastructure and applications has to be contextual.
Context binds people and things to a common scope; and hence, will ease mining of relevant information.
There has to be semantic knowledge that captures system and social behaviour, some specified while others are learned using models. 
Often, intelligent agents will act on the behalf of humans, and may be aware of personal preferences – Apple's Siri and Microsoft's Cortana are examples, and these will interact with digital agents of service providers, utilities and vendors. 
Semantic context will have to complement web standards for structural syntax to allow such M$2$M interaction to be effective.  

\subsection{Business Canvas}

If the IoT is to yield successful business models, we first need to recognize that IoT is not a new product or market. 
What IoT brings is an additional set of technologies, lower power, more computation and storage, cheaper devices, better wireless connectivity, much more granular control and observation capabilities. What it enables is scaling in both directions - up and down, and the ability to look at ourselves and the world in an unprecedented degree of detail.

IoT business models fall into two broad categories. 
Horizontals, concerned with enabling components and technology and verticals, which integrate these technologies to supply an end user with a value proposition. 

The first set of horizontal business models is the development of specific sensors and actuators that enable the generation of new, or more cost effective observations. 
The second is the deployment horizontal, a business model that addresses the needs of building out to scale of the data gathering, data storage and data curation and data brokering needs of IoT based systems. 
The third horizontal business model addresses the needs for a portfolio of analytical techniques to convert the data gathered into actionable information. 
While the first two have been the focus of IoT's precursor technologies, IoT's scale is driving active development across the board.

Verticals will pull solutions and services across these horizontals to deliver final end customer value. 
The emphasis here will be on the necessary domain and system integration expertise and the ability to build the necessary collaborations across customers and suppliers.

Building solutions at scale across these business models will require concerted efforts to support the modularization of the architecture, to provide access to capabilities through service based models, and of course real customer problems.

\bibliographystyle{unsrt}
\bibliography{IoT-extended-abstract} 

\end{document}